\documentclass[symmetry,article,submit,moreauthors,pdftex]{Definitions/mdpi}

\firstpage{1}
\pubvolume{1}
\issuenum{1}
\articlenumber{0}
\pubyear{2022}
\copyrightyear{2022}
\datereceived{}
\dateaccepted{}
\datepublished{}
\hreflink{https://doi.org/} % If needed use \linebreak
\Title{Paving the Way for a Multiplanetary Future for Humanity: How the Relativistic motor may enable eventually interplanetary travel}

\TitleCitation{Relativistic Motor and Interplanetary Travel}

\Author{Asher Yahalom$^{1,2}$\orcidA{}*}

\AuthorNames{Asher Yahalom}

\AuthorCitation{Yahalom, A.}
\address{%
$^{1}$ \quad Ariel University, Faulty of Engineering, Department of Electrical \& Electronic Engineering, Ariel 40700, Israel; asya@ariel.ac.il\\
$^{2}$ \quad Ariel University, Center for Astrophysics, Geophysics, and Space Sciences (AGASS),
Ariel 40700, Israel;}

\corres{Correspondence: asya@ariel.ac.il; Tel. +972-54-7740294: }

\abstract{Current spacecraft mass are mostly fuel, this is dictated by the lack of fueling stations in space and also by the Tsiolkovsky rocket equation which defines the mass ratio needed to escape earths gravity. The Tsiolkovsky rocket equation gives a relationship between the mass ratio and the final velocity in multiples of the exhaust speed, and dictates a high mass ration for current exhaust speeds. A relativistic motor exchanging momentum and energy with the electromagnetic field may mitigate such considerations, enabling efficient interplanetary travel. In this paper we will discuss the advantages and challenges of this novel mover for space transportation.}


\begin{document}
%%%%%%%%%%%%%%%%%%%%%%%%%%%%%%%%%%%%%%%%%%

\nolinenumbers

\newcommand{\beq} {\begin{equation}}
\newcommand{\enq} {\end{equation}}
\newcommand{\ber} {\begin {eqnarray}}
\newcommand{\enr} {\end {eqnarray}}
\newcommand{\eq} {equation}
\newcommand{\eqs} {equations }
\newcommand{\mn}  {{\mu \nu}}
\newcommand{\abp}  {{\alpha \beta}}
\newcommand{\ab}  {{\alpha \beta}}
\newcommand{\sn}  {{\sigma \nu}}
\newcommand{\rhm}  {{\rho \mu}}
\newcommand{\sr}  {{\sigma \rho}}
\newcommand{\bh}  {{\bar h}}
\newcommand{\br}  {{\bar r}}
\newcommand {\er}[1] {equation (\ref{#1}) }
\newcommand {\ern}[1] {equation (\ref{#1})}
\newcommand {\Ern}[1] {Equation (\ref{#1})}
\newcommand{\hdz}  {\frac{1}{2} \Delta z}
\newcommand{\curl}[1]{\vec{\nabla} \times #1} % for curl
\newcommand {\Sc} {Schr\"{o}dinger}
\newcommand {\SE} {Schr\"{o}dinger equation }
\newcommand{\ce}  {continuity equation }

\section{Introduction}

Today's space vehicles total mass are mostly fuel. The reason for this is simple there are no fueling  stations in space. This problem will become even more severe as interplanetary travel is envisioned. 

Motion means kinetic energy and momentum. While energy is abundant in interplanetary space as sunlight can be converted to other forms of energy, the real problem is momentum. Since momentum
is conserved, how can one entail motion which means momentum creation? This problem is solved on earth by the car gaining momentum by pushing the road backward. Similarly a jet plane is propagating
by pushing the air behind it. That is forward momenta is gained by the vehicle while at the same time generating momenta of the same magnitude but in opposite direction in the surrounding medium. 
The total momentum gained is of course null, which is the essence of linear momentum conservation.

But what can one push against in empty space? The common answer is against nothing. So how can we travel? Again the common wisdom is by the rocket mechanism that is carrying with us material and ejecting it as we go. The momentum gained by the vehicle is equal but opposite in sign to the momentum of the ejected material. This situation dictates that a huge part of a spacecraft devoted to an interplanetary mission must be just fuel.

The situation becomes even more dire when one needs to take into account that a space craft 
must escape a deep gravitational well. First the Earths gravitational well and then the gravitational well of the planet to which it travels (Mars?) when coming back. 

\begin{figure}[H]
\centering
\includegraphics[width=0.7\columnwidth]{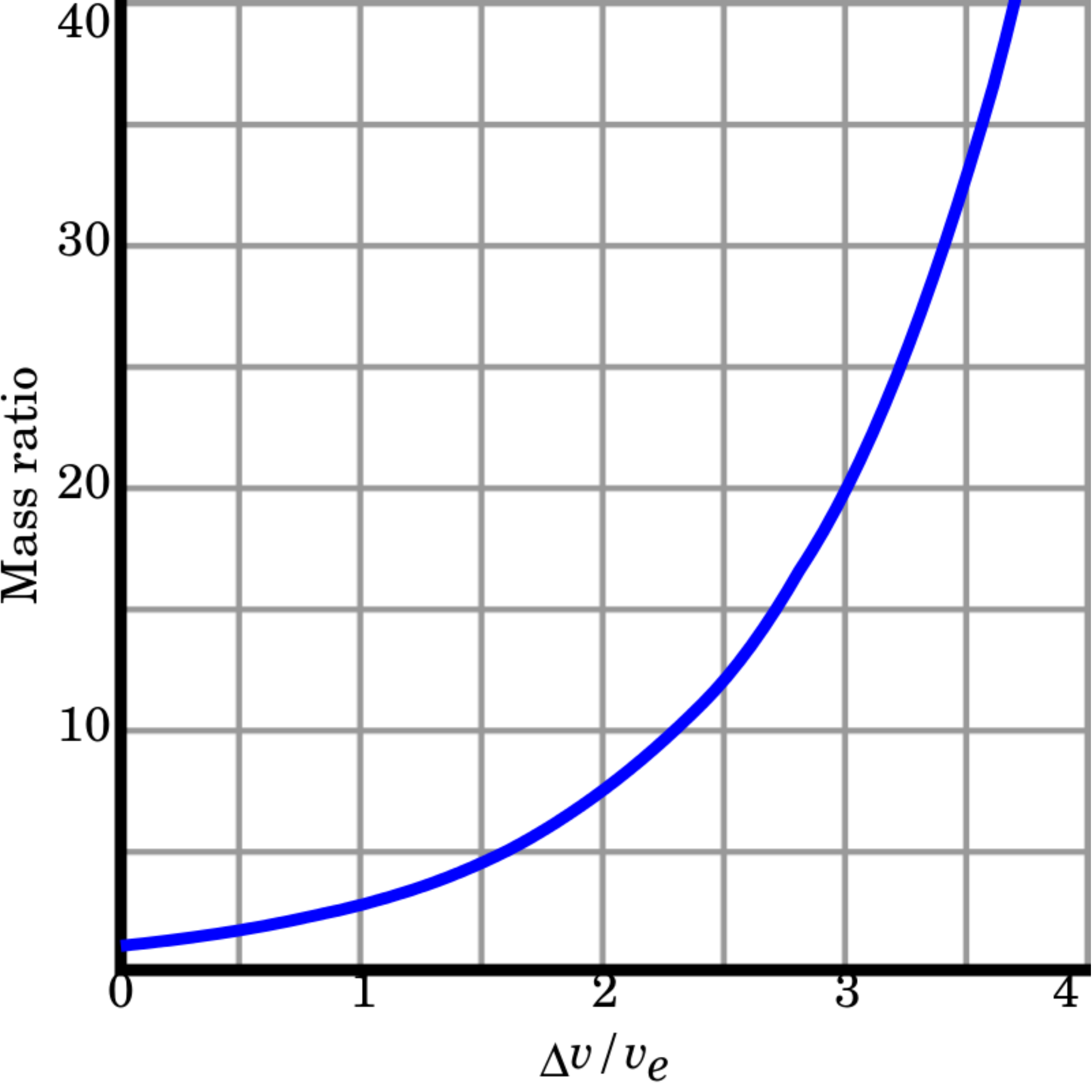}
\caption {A rocket's required mass ratio as a function of effective exhaust velocity ratio.}
 \label{Tsiolkovskyf}
\end{figure}

The Tsiolkovsky rocket equation \cite{Tsil} defines the mass ratio needed to escape earths gravity. The Mass ratio is the ratio between the rocket's initial mass and its final mass. The Tsiolkovsky rocket equation gives a relationship between the mass ratio and the final velocity in multiples of the exhaust speed, and dictates a high mass ratio for current exhaust speeds.
\beq
\Delta v = v_e \ln \left( \frac{m_0}{m_f}\right)
\label{Tsiolkovsky}
\enq 
in the above $\Delta v$ is the maximum change of velocity of the vehicle (with no external forces acting), $v_e$ is the effective exhaust velocity, $m_0$ is the initial total mass that is including the propellant mass, and $m_f$ is the final total mass without propellant.
The equation is depicted in figure \ref{Tsiolkovskyf}.
As typical exhaust velocities are \cite{Sutton}:
\begin{itemize}
\item 1.7 to 2.9 km/s for liquid monopropellants.
\item 2.9 to 4.5 km/s for liquid bipropellants.
\item 2.1 to 3.2 km/s  for solid propellants.
\end{itemize}
and the earth escape velocity is 11.2 km/s, it follows from figure \ref{Tsiolkovskyf} that a large mass ratio is needed just to escape the earth's gravity not to mention reaching an appreciable velocity that will allow a reasonable travel time to a nearby planet. 

The above considerations seem unescapable. Indeed,  without propellant how can one hope to defeat the requirement to conserve linear momentum and with it the required mass ratio? And yet one must not forget the linear momentum is not only a property of matter but also a property of the electromagnetic field \cite{Jackson,Feynman,MTAY4}. Thus in principle a space vehicle might 
propagate using the energy supplied by the sun and contained within its storage devices while 
the momentum it gains is balanced by the same amount of momentum but of opposite direction which
is transferred to the electromagnetic field.

A relativistic motor exchanging momentum and energy with the electromagnetic field may mitigate mass ratio considerations, enabling efficient interplanetary travel. In this paper we will discuss the advantages and challenges of this novel mover for space transportation.

A detailed introduction to the subject of relativistic motors in general and microscopic relativistic engines in particular with suitable references can be found in \cite{nano} and will not repeated here, the interested reader is referred to the original text. A brief history of the relativistic engine is given below.

\begin{figure}
\centering
\includegraphics[width=0.7\columnwidth]{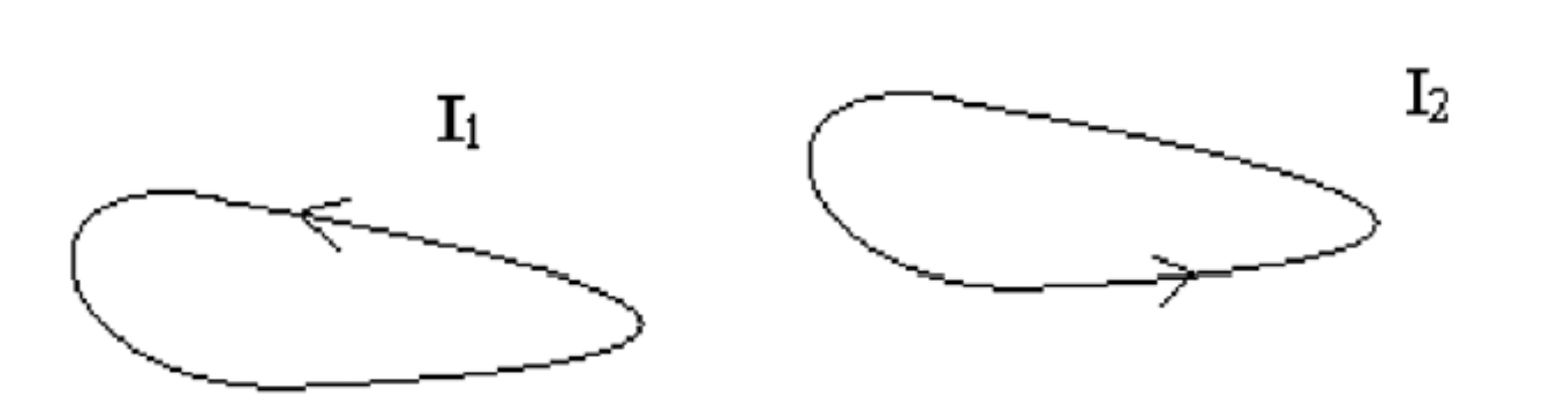}
\caption {Two current loops.}
 \label{twoloops}
\end{figure}

The first relativistic engine suggested was base on electromagnetic field retardation
of two time dependent loop currents \cite{MTAY1} (see figure \ref{twoloops}). 
Jefimenko's \cite{Jefimenko,Jackson} formula was used to calculate the total force $\vec F_{T}$ operating on the center of mass of a system resulting in the formula:
\beq
\vec F_{T} \cong \frac{\mu_0}{8 \pi} (\frac{h}{c})^2  \vec K_{122}  I_2  I_1^{(2)}(t),
\qquad I_1^{(n)}(t) \equiv  \frac{\partial^n I_1 (t)}{\partial t^n}
\label{Ftotal4}
\enq
in which $\mu_0$ is the magnetic permeability of the vacuum, $c$ is the velocity of light in the vacuum, $h$ is at typical length scale of the system and $\vec K_{122}$ is a dimensionless vector which depend on the geometry of the loops. $I_2$ is a static current and $I_1 (t)$ is a time dependent current. This was later generalized to calculate the total force in a system of permanent magnet and a current loop \cite{AY1}. As force is applied for a finite duration,  momentum will be acquired and kinetic energy for the entire system. It may superficially seem that the laws of momentum and energy conservation are violated, but this is not so. Linear momentum conservation was validated in \cite{MTAY4}. It was shown that the momentum gained by the field $\vec P_{field~12} $ is the same as momentum gained by the engine $\vec P_{mech}$ but in an opposite direction:
\beq
\vec P_{field~12} = - \frac{\mu_0}{8 \pi} I_2  I_1^{(1)} (t) \frac{h^2}{c^2} \vec K_{122}
 = -\vec P_{mech}.
\label{Pf12h}
\enq
The exchange of energy between the kinetic part of the relativistic engine and the electromagnetic field was elaborated in \cite{AY2,RY2,RY3}. It was demonstrated that the electromagnetic energy consumed is six times the kinetic energy provided to the engine. It was also shown that energy is radiated from the engine if the coils are misaligned.

Our preliminary analysis assumed bodies that were electric charge natural. In a later paper \cite{RY4} charged bodies  were analyzed. The charged engine allows to maintain a finite momentum even if the current is not continuously increasing, as is dictated by the current derivative term
in \ern{Pf12h} (which requires a monotonously increasing current for a uniform motion in some direction). This more general case result is a total force in the center of mass and total linear momentum given by the formulae:
\beq
\vec F_{T}= \frac{\mu_0}{4 \pi} \partial_t \int \int d^3 x_1  d^3 x_2 ~
  \left[\frac{1}{2}\left(\rho_2 \partial_t \rho_1 - \rho_1 \partial_t \rho_2 \right)\hat R - (\rho_1  \vec J_2 + \rho_2  \vec J_1)  R^{-1}  \right], \qquad \vec R \equiv \vec x_1 - \vec x_2
\label{F212t3}
\enq
\beq
\vec P_{mech} (t)= \frac{\mu_0}{4 \pi}\int \int d^3 x_1  d^3 x_2 ~
  \left[\frac{1}{2}\left(\rho_2 \partial_t \rho_1 - \rho_1 \partial_t \rho_2 \right)\hat R - (\rho_1  \vec J_2 + \rho_2  \vec J_1)  R^{-1}  \right]
\label{P2}
\enq
in the above $\rho$ is the charge density and $\vec J$ is the current density of the two
subsystems 1 and 2 respectively, and an integrations is required over the volumes of the two sub
systems. 

\begin{figure}[H]
\centering
\includegraphics[width=\columnwidth]{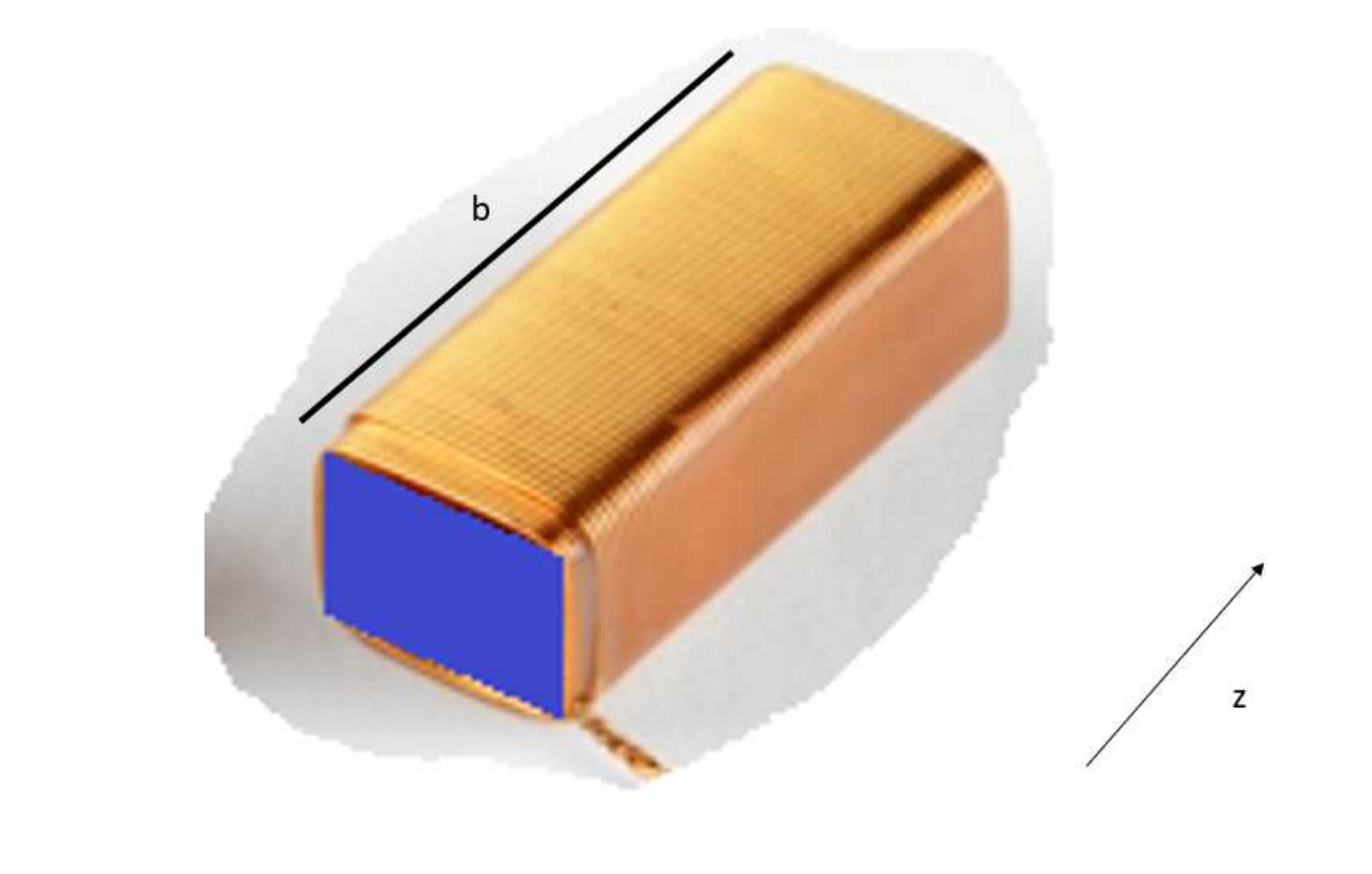}
\caption {A relativistic engine.}
 \label{releng1}
\end{figure}

\begin{figure}[H]
\centering
\includegraphics[width=\columnwidth]{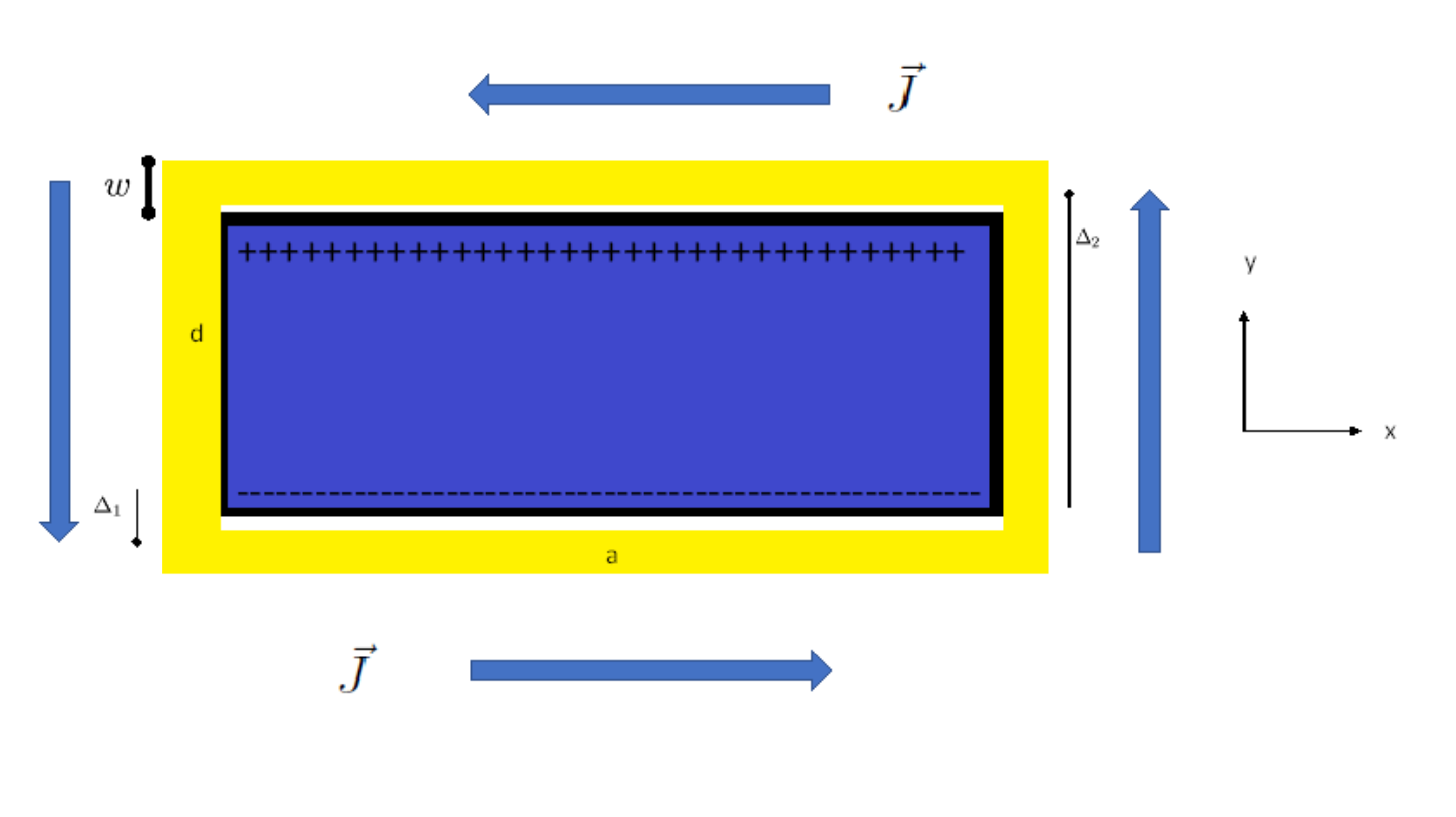}
\caption {A cross section of the relativistic engine.}
 \label{releng2}
\end{figure}

However, due to dielectric breakdown which dictated a maximal value to charge density and current density limitations that can be transferred even through a superconducting wire it is shown that for any reasonable geometrical size the momentum that can be gained in a relativistic charged engine is rather limited. Table \ref{partabb} obtained in \cite{RY4} demonstrates the severe limitations of macroscopic configurations:
\begin{table}
\begin{center}
\begin{tabular}{|c|c|c|c|c|}
  \hline
           &car &  rocket size engine & giant cube & units \\
  \hline
  $a$ & 6 & 200 & 1000 & m \\
  $b$ & 2 & 10 & 1000 & m \\
  $d$ & 1 & 10 & 1000 & m \\
  $w$ & 0.2 & 0.4 & 0.4 & m \\
  \hline
  $P_{mech}$ & 0.3 & 868 & $3.1 \ 10^7$  & kg m/s \\
  \hline
\end{tabular}
\end{center}
\caption{Maximal momentum gained by a relativistic motor for three cases of parameters. We assume an extreme case of charge density $\sigma = 3.7 \ 10^ {-3}$ Coulomb/$\rm m^2$, and current density $J_0 = 5 \ 10^{7}$  Ampere/$\rm m^2$.}
\label{partabb}
\end{table}
The physical structure of this particular relativistic engine and its geometric parameters are depicted in figure \ref{releng1} and figure \ref{releng2}. 

The above limitations suggested to use the high charge densities that are available in the microscopic realm, for example in ionic crystals. We have pursued this idea in a previous paper
\cite{nano} in which we calculated the high charge densities and current densities in the atomic level. We also deduced a preliminary form for the optimized wave function in terms of relativistic engine performance in two cases a wave packet in a hydrogen atom and a eigen state in a simple molecule which introduces a static electric field of broken spherical symmetry.

To conclude the relativistic motor is well suited for space travel and interplanetary motion 
as it posses the following attributes:
\begin{itemize}
\item Allows 3-axis motion (including vertical)
\item No moving parts
\item Zero fuel consumption
\item Zero carbon emission
\item Needs only electromagnetic energy (which may be provided by solar panels).
\item Ideal  solution for space travel in which currently much of the space vehicle mass is devoted to fuel
\item Highly efficient, in principle kinetic energy can be converted back to electromagnetic energy.
\end{itemize}

However, to reach a practical relativistic engine that will serve humanity in interplanetary travel one must manipulate matter at subatomic levels, a feat that is quite challenging. In this paper we shall investigate two ways of doing so one that is related to free electrons and the other to confined electrons. While we start with a classical description of the problem we cannot and do not
ignore the fact that on the atomic level a quantum description is required. It will be shown that
the quantum effects are much more important for confined electrons than for free electrons.  
We shall not derive the basic equations of the relativistic engine here the
interested reader is referred to \cite{RY4,nano}, we shall also use the same notations as in the previous papers and will not redefine the symbols.

In the current paper we will show that both free electrons and confined electrons can be put in a configuration supporting a relativistic motor effect which will allow eventually the construction of interplanetary space-craft. However, quantum mechanics (for spin and spin less electrons) is only important in the confined electron case. We thus derive the form of the electromagnetic field needed to maintain the appropriate wave packet that support a relativistic engine effect in the confined case.

\section{Relativistic engine in the microscopic scale}

\subsection{A classical electron}

Before introducing quantum considerations we shall first consider a classical system of two point particles each with a charge of absolute value $|e|$. We shall assume one charge to be stationary
while the other moving with velocity $\vec v_2$, it thus follows that system $2$ has current density
of \cite{RY4}:
\beq
\vec J_2 = \rho_{2} \vec v_{2} = e \delta^3 (\vec x_2 - \vec x (t)) \vec v_2
\label{pcurden}
\enq
and system $1$ has a charge density:
\beq
\rho_{1} =  \pm e \delta^3 (\vec x_1),
\label{chden}
\enq
in which we assume for convenience that the stationary charge is located at the origin of the coordinates.

\subsubsection{Proximity considerations}

\begin{figure}[H]
\centering
\includegraphics[width=0.7\columnwidth]{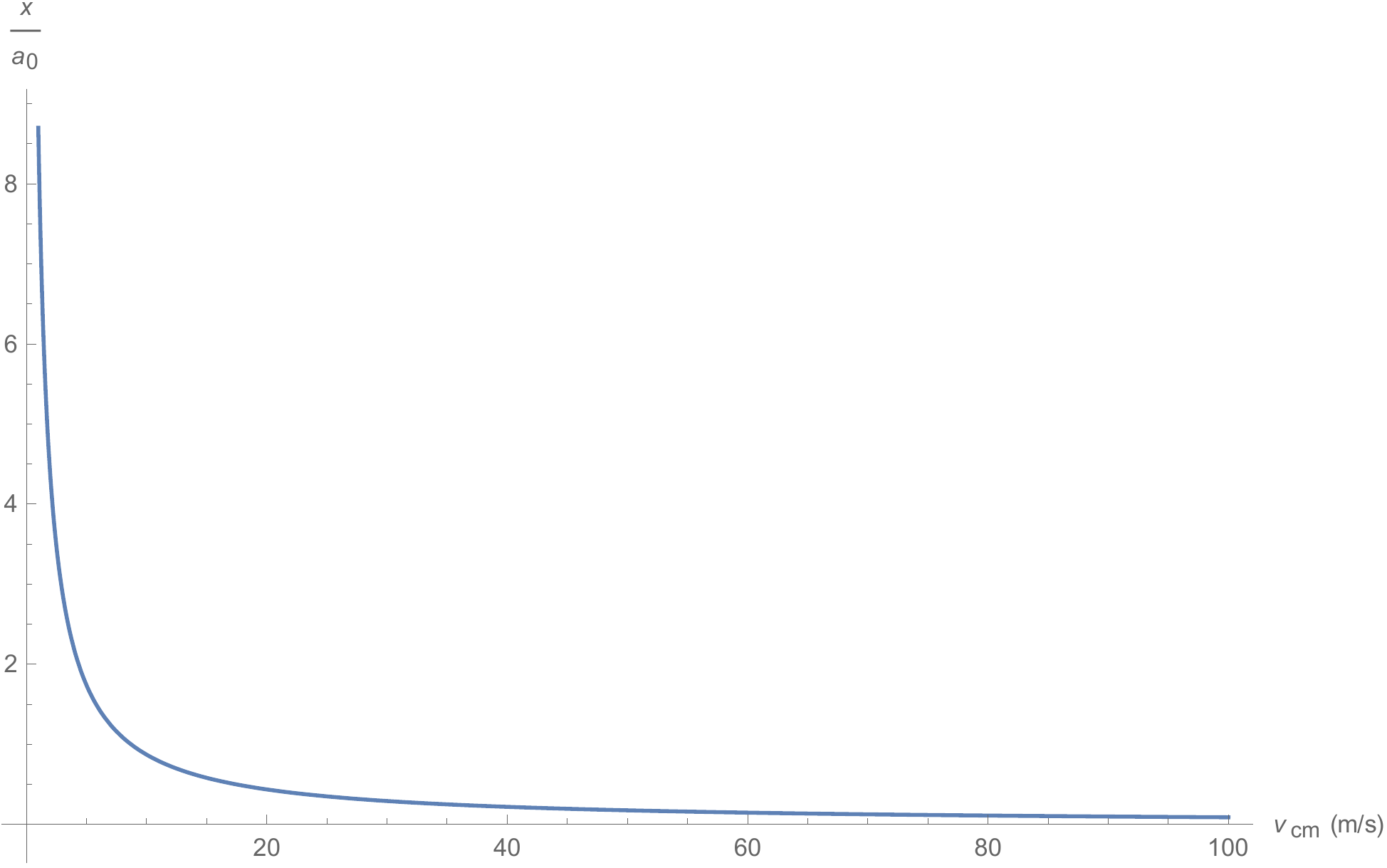}
\caption {The proximity between a classical electron and proton, needed to achieve a desired velocity for an unloaded engine.}
 \label{Xdf}
\end{figure}

Plugging \ern{pcurden} and \ern{chden} into equation (3) of \cite{nano}:
\beq
\vec P (t)= - \frac{\mu_0}{4 \pi} \int \int d^3 x_1  d^3 x_2 ~ \rho_1  \vec J_2  R^{-1} =
= \mp \frac{\mu_0 e^2}{4 \pi} \frac{\vec v_2}{|\vec x (t)|}
\label{P42pp}
\enq
Taking the total mass of the two particle system to be $m_t = m_1 + m_2$ we arrive at a center of
mass velocity:
\beq
\vec v_{cm} (t) = \frac{\vec P (t)}{m_t}= \mp \frac{\mu_0 e^2 }{4 \pi m_t} \frac{\vec v_2}{|\vec x (t)|}.
\label{vcm}
\enq
The above equation makes explicit the fundamental conflicting requirements of the concept. To have significant speed in the center of mass the particles must be close to each other thus we would like to have a confined system. On the other hand we would like to have a high $v_2$ with a constant direction, this is impossible in a confined system as in such a case, $ \vec v_2$ must eventually change direction. Thus for the center of mass to obtain a speed $v_{cm} (t)$ at time $t$ the particles must be at the proximity:
\beq
|\vec x (t)|= \mp \frac{\mu_0 e^2 }{4 \pi m_t} \frac{v_2}{v_{cm} (t)}.
\label{prox1}
\enq
If the particles are an electron and a proton $m_t \simeq m_p$
\beq
|\vec x (t)|= \frac{\mu_0 e^2 }{4 \pi m_p} \frac{v_2}{v_{cm} (t)}.
\label{prox2}
\enq
It is not difficult to bring an electron to move very close to the speed of light
such that: $v_2 \simeq c$, for example for a 99\% of the speed of light with a low energy accelerator:
\beq
v_2 = 0.99 c \quad \Rightarrow \quad E_k = \frac{m_e c^2}{\sqrt{1-(\frac{v_2}{c})^2}}
\simeq 3.6 {\rm MeV}.
\label{Ekspl}
\enq
thus we shall take $v_2 = c$ and obtain:
\beq
|\vec x (t)|=  \frac{\mu_0 e^2 c }{4 \pi m_p} \frac{1}{v_{cm} (t)} = \frac{4.6 \cdot 10^{-10}}{v_{cm} (t)} = 8.7 \frac{a_0}{v_{cm} (t)}.
\label{prox3}
\enq
in which the last expression contains the Bohr radius:
\beq
a_0 \equiv \frac{4 \pi \epsilon_0 \hbar^2}{m'_e e^2} \simeq \frac{\hbar^2}{k m_e e^2} \simeq 0.53~10^{-10} {\rm ~ m}~, \qquad
m'_e \equiv \frac{m_e m_p}{m_e + m_p} \simeq m_e.
\label{Bohrrad}
\enq
The relation between the require distance and the desired velocity for the two particle system is described in figure \ref{Xdf}.
 Thus a typical car's velocity: $v_{cm} = 50$ m/s $=180$ km/h is obtained for:
 \beq
|\vec x (t)|  \simeq 0.174~ a_0.
\label{x50}
 \enq
 If we require the hydrogen relativistic engine to reach the earth's escape velocity of $v_{cm} = 11.2$ km/s then we must have a proximity of:
 \beq
|\vec x (t)| ~ \simeq 48.8 ~ r_p
\label{xescvel}
 \enq
in which $r_p = 8.4 ~10^{-16} $ m is the proton charge radius. Thus the
distance between electron and proton must be of a nuclear size rather than an atomic size. Finally if we imagine that the relativistic engine will reach relativistic velocities $v_{cm} \simeq 0.1 c$ it follows that:
\beq
|\vec x (t)| \simeq 0.018 ~ r_p
\label{xlightspeed}
 \enq
that is the electron proton system is of sub nuclear dimensions. 

An engine suitable for interplanetary travel must include a macroscopic amount of such atoms and it must carry not just itself but also some payload.

\subsubsection{An unconfined electron}

\begin{figure}[H]
\centering
\includegraphics[width=0.7\columnwidth]{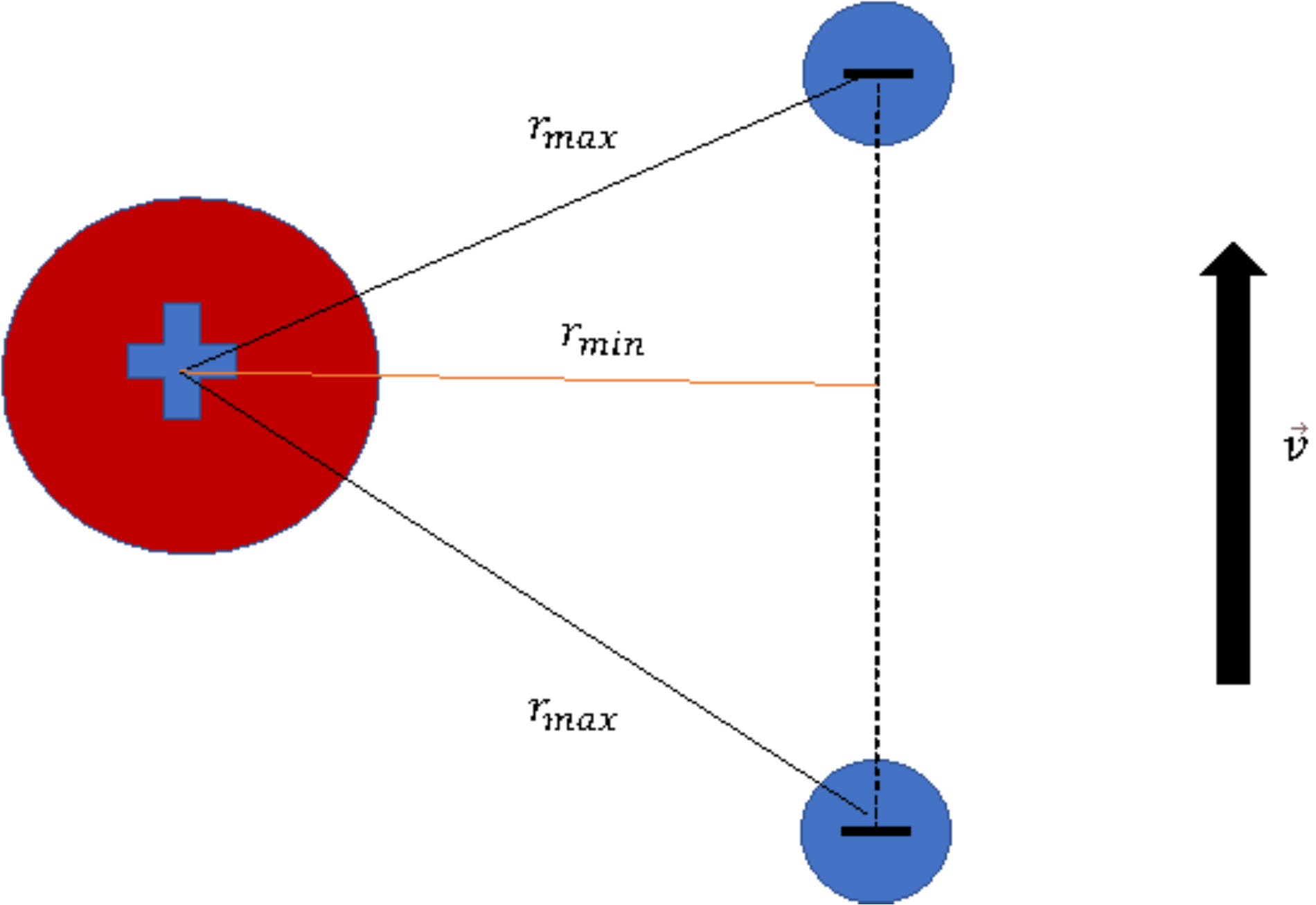}
\caption {Two electrons from a train of electrons, moving in the vicinity of a proton.}
 \label{etrf}
\end{figure}

\begin{figure}[H]
\centering
\includegraphics[width=0.4\columnwidth]{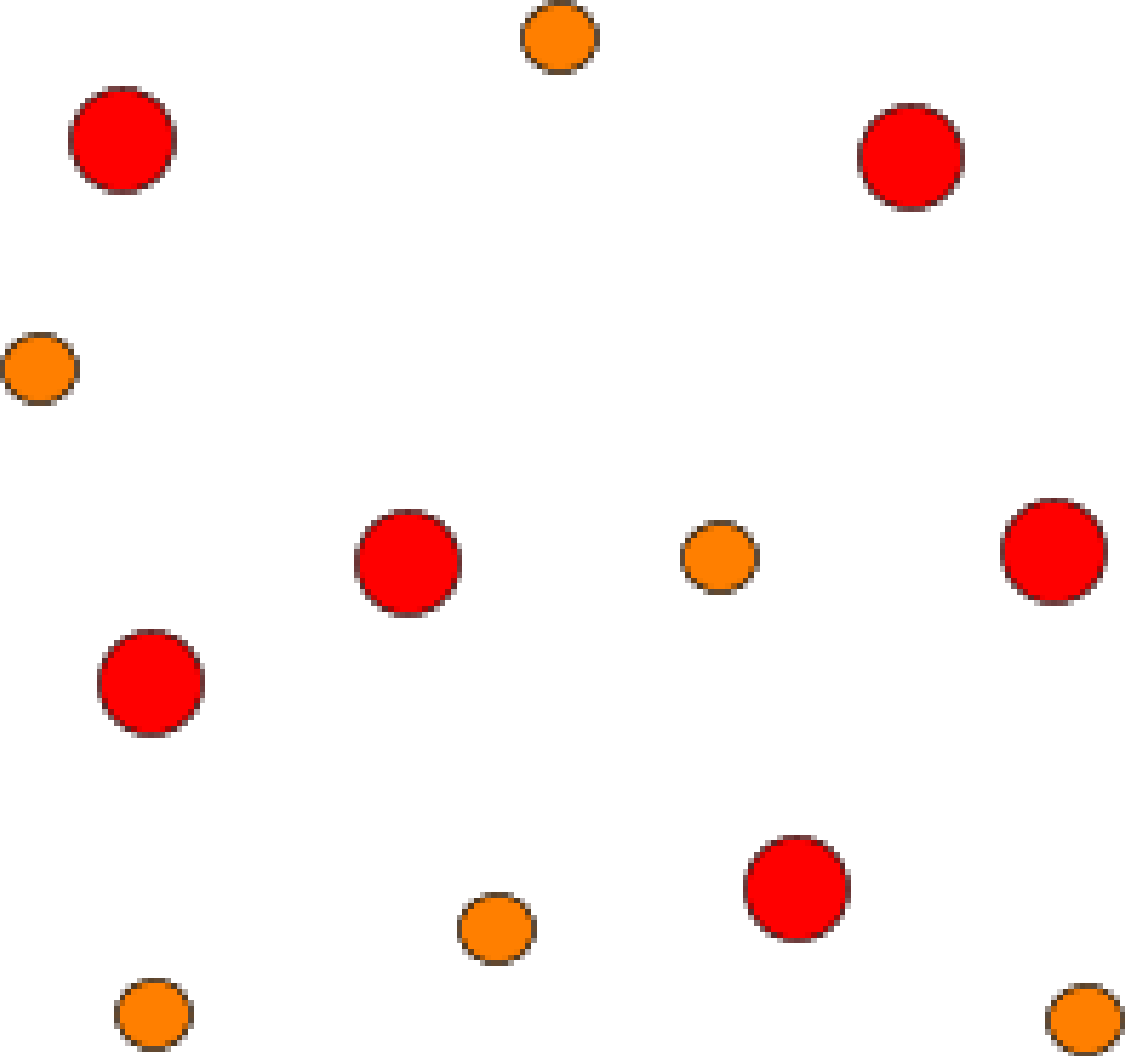}
\caption {Plasma of protons and electrons, the read circles symbolize protons, while the green circles symbolize electrons.}
 \label{plasmaf}
\end{figure}

It seems that a way to circumvent this inherent contradiction between proximity and velocity is to use a train of particles in which for each particle leaving the desired range a new one enters as depicted in figure \ref{etrf}.
Let us assume that we keep at least one electron from the proton at a distance which is not
smaller than a distance $r_{max}$ and bigger that a distance $r_{min}$, it follow that
the duration between successive electrons is:
\beq
\Delta t = 2 \frac{\sqrt{r_{max}^2 - r_{min}^2}}{c}
\label{delt}
\enq
If we take $r_{min}$ to be the distance for a $v_{cm} = 50$ m/s, that is  $r_{min} = 0.174~ a_0$
and $r_{max}=3 r_{min}$
it follows that:
\beq
\Delta t \simeq 1.7 \cdot 10^{-19} {\rm s} \quad \Rightarrow \quad I = \frac{e}{\Delta t} \simeq
0.92 ~{\rm A}
\label{delt2}
\enq
hence the needed current is not too excessive. The practical problem is how to put high velocity electrons in the vicinity of protons. One may imagine, a high density plasma (see figure \ref{plasmaf})
made of protons and electrons in which the average density between protons is $2 r_{min}$ the density of such a plasma is:
\beq
\Delta V = (2 r_{min})^3  \simeq 6.2 \cdot 10^{-33} {\rm m^3} \quad \Rightarrow \quad \rho_{plasma} = \frac{m_p}{\Delta V} \simeq 2.7 \cdot 10^5 ~{\rm Kg/m^3}
\label{plasmadensity}
\enq
this is much higher than the density of solid hydrogen $rho_{SH} \simeq 86 ~{\rm Kg/m^3}$ in which the typical distance between atoms is about $5$ Bohr radii. We notice that a $3.7$ Liter of the above plasma will weigh about one metric ton, and could easily move a standard car. We also notice that, solid hydrogen can only be obtained under unusual conditions of low temperature and high pressure. Another alternative is to have sparse protons but high density electrons with typical distance of $2 r_{min}$. However, this will lead to for the $\Delta V$ of \ern{plasmadensity} to a charge density of:
\beq
\rho_{charge} = \frac{e}{\Delta V} \simeq 2.6 \cdot 10^{13} ~{\rm C/m^3}
\label{echargedensity}
\enq
Obviously, such a configuration cannot hold. If we take for simplicity the configuration to
be spherical of radius $r_s$, then according to equation (9) of \cite{nano} the electric field on its surface would be:
\beq
E_r = \frac{k Q}{r_s^2} = \frac{k \frac{4 \pi}{3} \rho_{charge} r_s^3}{r_s^2}
= k \frac{4 \pi}{3} \rho_{charge} r_s  \simeq 9.7 \cdot 10^{23} r_s
\label{Eradial}
\enq
Thus an electron on the surface of the said sphere will be accelerated outwards with
an acceleration of:
\beq
\ddot{r} =  \frac{e E_r}{m_e}  \simeq 1.7 \cdot 10^{35} r_s~ {\rm m/s^2}.
\label{outwardaccelr}
\enq
The typical disintegration time of the above configuration is:
\beq
\tau_{disintegration} = \sqrt{\frac{r_s}{\ddot{r} }} \simeq  2.4 \cdot 10^{-18}~ {\rm s}.
 \label{disintegration}
\enq
regardless of the size of the sphere. We shall make a point regarding the typical charge separation that is empirically available. According to section \ref{chden} the maximal charge density for air is $\sigma_{max} \simeq 53 ~ \mu C /m^2$. In terms of electron number density this is:
\beq
\sigma_{max ~ electrons} = \frac{\sigma_{max}}{|e|} \simeq 3.3 \cdot 10^{14} ~{\rm m^{-2}}
\label{smaxelec}
\enq
which translates into a typical spatial separation of:
\beq
\delta_e = \frac{1}{\sqrt{\sigma_{max ~ electrons}}} \simeq 5.5 \cdot 10^{-8} ~{\rm m}
\label{dele}
\enq
this separation is much too large to obtain a significant relativistic motor effect.
On the other hand, looking back at the ionic crystal of figure 1 of \cite{nano} it is easy to draw a trajectory for the said stream of electrons as depicted in figure \ref{Nacltraj}:
\begin{figure}[H]
\centering
\includegraphics[width=\columnwidth]{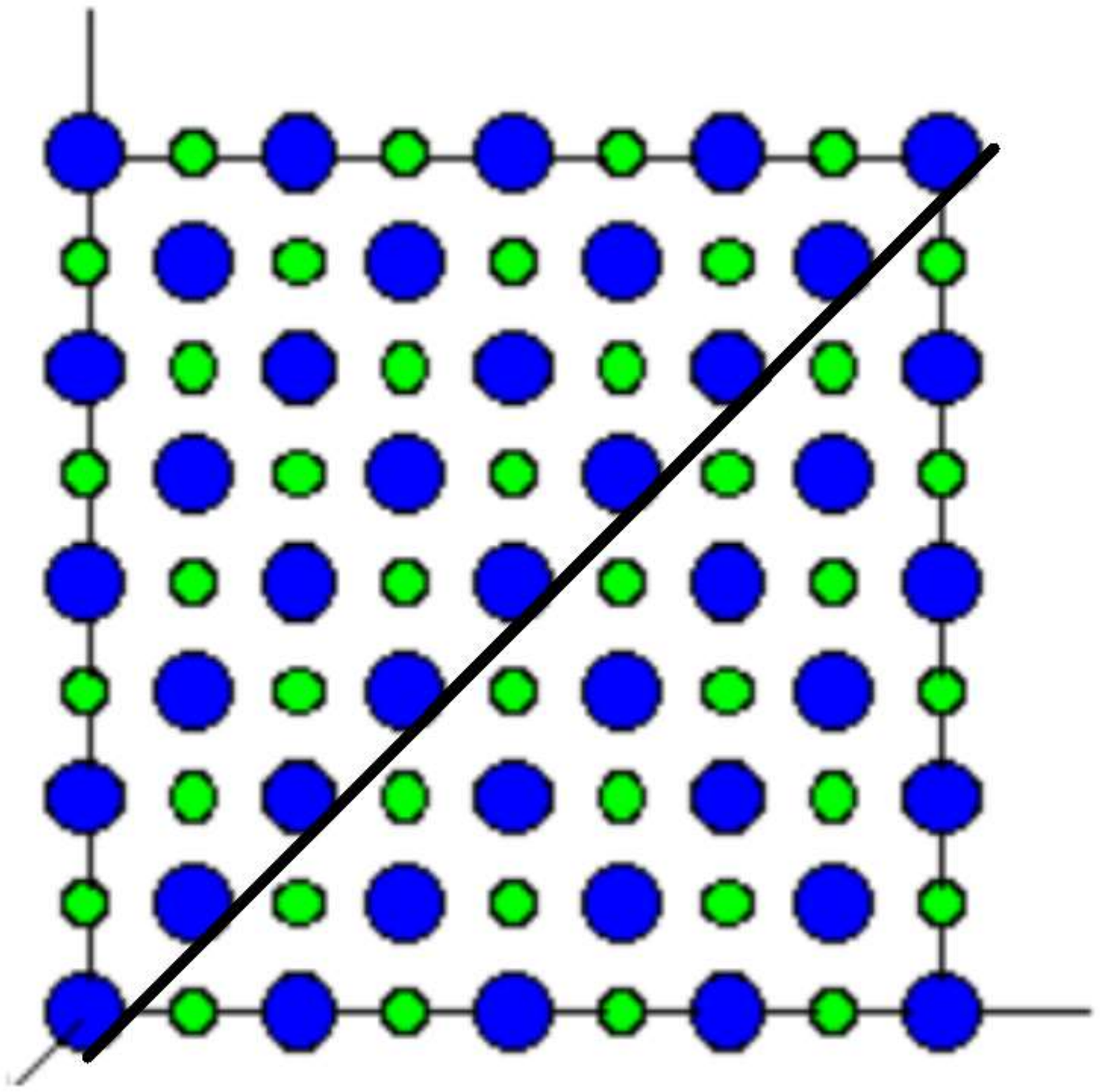}
\caption {Depicted is the $100$ plane of a lattice of table salt $Na^+ Cl^-$, blue circles are Sodium positive ions and yellow circles are Chlorine negative ions. The trajectory of relativistic electrons is described in terms of thick black line.}
 \label{Nacltraj}
\end{figure}
As the electron passes through more positive ions it becomes closer to the positive ion, if fact
even an electron at a distance of a lattice constant of $ l \simeq 564$ pm will feel a force perpendicular to its trajectory and towards the positive ions line of about:
\beq
F_{\perp} \simeq \frac{k e^2}{l} \simeq 4 \cdot 10^{-19}~ {\rm Newton}.
 \label{Fperp}
\enq
Thus it will have a perpendicular acceleration towards the positive ionic line of about:
\beq
\ddot{r}_{\perp} = \frac{F_\perp}{m_e} \simeq 4.5 \cdot 10^{11}~ {\rm m/s^2},
 \label{accperp}
\enq
for a duration of about $\Delta t$ (see \ern{delt2}), in each ion passage. Thus the velocity
towards the positive ion line would be at least:
\beq
v_{\perp} \simeq  \Delta t \ddot{r}_{\perp} \simeq 7.8 \cdot 10^{-8}~ {\rm m/s},
 \label{vperp}
\enq
but of course the acceleration and velocity will become larger as the electron reaches closer to
the ion line. Thus the electron will reach the ion line in a time shorter than:
\beq
 \tau_{ion~line} \simeq \frac{v_{\perp}}{l} \simeq 0.007 ~ {\rm s},
 \label{tioline}
\enq
This time can be shortened by applying an external electric field perpendicular to the ion line and away from the line. Moreover, a slower electron beam will have a more time to converge to the ion line which poses an interesting optimization problem, balancing between the desired proximity to the ion line and the electron beam speed. We notice that a $99 \%$ speed of light will hardly converge
to the ion line even if the engine is one meter thick, because it will pass it in about three nano seconds. If convergence to the ion line is indeed achieved we expect an oscillatory motion around the positive ion line in which inverse beta decay will occasionally occur. Of course the most significant relativistic motor effects will occur in the times in which the electron is closer to the positive ion line.

\subsubsection{A confined electron}

\begin{figure}[H]
\centering
\includegraphics[width=0.7 \columnwidth]{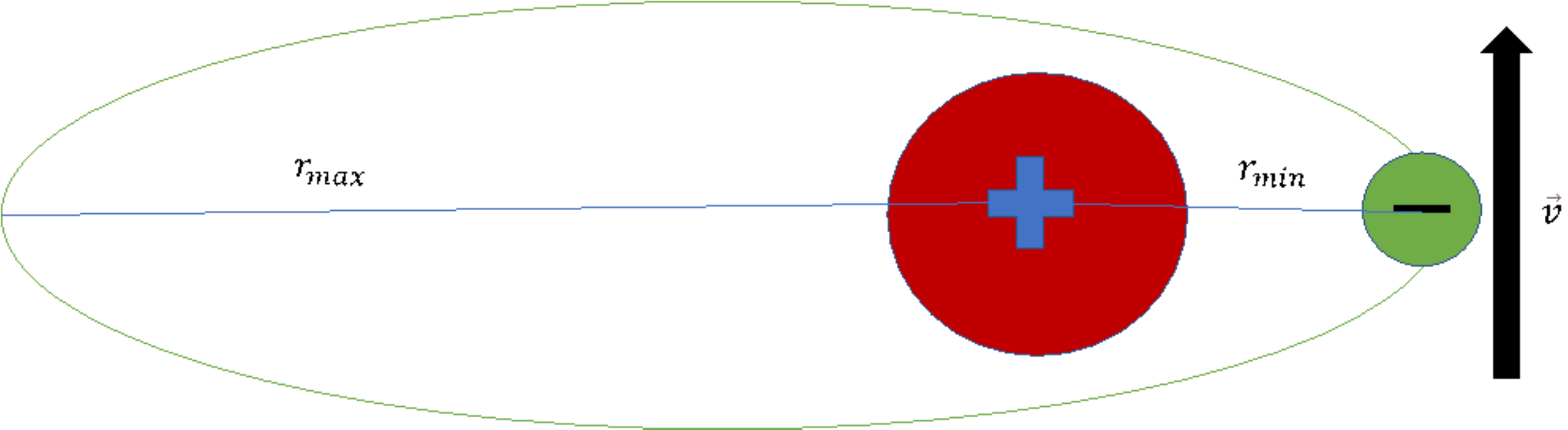}
\caption {A schematic of an elliptical orbit of an electron around a proton.}
 \label{Elliptic}
\end{figure}

A confined classical electron can be put in an elliptical trajectory that can be occasionally favourable to the relativistic motor effect, see figure \ref{Elliptic}.
One can see that the positive relativistic motor effect near the proton at a distance $r_{min}$, is much greater than the negative relativistic motor effect which occurs due to the motion in the opposite direction but at a much larger separation $r_{max}$. We also notice that the orthogonal motions will cancel each other as they occur in opposite directions. Unfortunately such a description is not very useful as quantum effects play a major rule for confined electron, thus we conclude our discussion regarding classical electrons and move to the discussion of quantum electrons.

\section{Schr\"{o}dinger's electron}

Quantum mechanics according to the Copenhagen interpretation has lost faith in our ability to predict precisely the whereabouts of even a single particle. What the theory does predict precisely is the evolution in time of a quantity
denoted "the quantum wave function", which is related to a quantum particle whereabouts in a statistical manner. This evolution is described by an equation suggested by
Schr\"{o}dinger \cite{Schrodinger}:
\beq
i \hbar \dot{\psi} = \hat{H}_S \psi, \qquad \hat{H}_S = -\frac{1}{2m_e}
 \left(\hbar \vec \nabla - i e \vec A\right)^2 - e \Phi
 \label{Seq}
 \enq
in the above $i=\sqrt{-1}$ and $\psi$ is the complex wave function. $\dot{\psi}=\frac{\partial \psi }{\partial t}$ is the partial time
derivative of the wave function. $\hbar=\frac{h}{2 \pi}$ is Planck's constant divided by $2 \pi$.
 However, this presentation of quantum mechanics is
 rather abstract and does not give any physical picture regarding the meaning of the quantities involved. Thus we write the quantum wave function using its modulus $a$ and phase $\phi$:
 \beq
\psi = a e^{i\phi}.
%\label{nrpsi}
\label{psi1}
\enq
The probability density and flux are defined as:
\beq
\tilde \rho = \psi^* \psi, \qquad \vec J_S =  \frac{\hbar}{2 m_e i} [\psi^* \vec \nabla \psi -(\vec \nabla \psi^*) \psi] - \frac{e}{m_e} \vec{A} \tilde \rho
 = \tilde \rho (\frac{\hbar}{m_e} \vec \nabla \phi - \frac{e}{m_e} \vec A).
\label{scd}
\enq
We thus define the velocity field using the natural definition:
\beq
\vec v_S = \frac{\vec J_S}{\tilde \rho}= \frac{\hbar}{m_e} \vec \nabla \phi - \frac{e}{m_e} \vec A
\label{vS}
\enq
 and the mass density is defined as:
 \beq
 \hat \rho= m_e \tilde \rho = m_e a^2.
 \label{massden}
 \enq
It is easy to show from \ern{Seq} that the \ce is satisfied:
 \beq
 \frac{ \partial\hat \rho }{ \partial t} + \vec \nabla \cdot (\hat \rho \vec v_S) = 0
 \label{nrec}
 \enq
  Hence $\vec v_S$ field is the velocity associated with  mass conservation.
However, it is also the mass associate with probability $a^2$ (by Born's
 interpretational postulate) and charge density $\rho= e a^2$.
 The equation for the phase $\phi$ derived from \ern{Seq} is as follows:
 \beq
  \hbar \frac{\partial \phi}{\partial t} +
  \frac{1}{2 m_e}\left(\hbar \vec \nabla \phi -e \vec A\right)^2 -  e \Phi =
   \frac{\hbar^2 \nabla^2 a }{2 m_e a} = -Q
 \label{nrhje}
 \enq
In term of the velocity defined in \ern{vS} one obtains the following equation of motion
(see Madelung \cite{Madelung} and Holland \cite{Holland}):
\beq
\frac{d \vec v_S}{d t}  = \frac{\partial \vec v_S}{\partial t} + (\vec v_S \cdot \vec \nabla) \vec v_S
= - \vec \nabla \frac{Q}{m_e}+\frac{e}{m_e} (\vec E + \vec v \times \vec B)
\label{EulerS}
\enq
 The right hand side of the above equation contains the "quantum correction":
 \beq
Q = -\frac{\hbar^2}{2 m_e} \frac{\vec{\nabla}^2 \sqrt{\hat \rho}}{\sqrt{\hat \rho}}.
\label{qupo}
\enq
For the meaning of this correction in terms of information theory see:
\cite{Spflu,Fisherspin,Fisherspin2}.  These results  illustrates the advantages
 of using the two variables, phase and modulus, to obtain equations of motion
that have a  substantially different form than the familiar \SE (although having the same
mathematical content) and have straightforward physical interpretations \cite{Bohm}.

The quantum correction $Q$ will of course disappear in the classical limit $\hbar \rightarrow 0$, but even if one intends to consider the quantum equation in its full rigor, one needs to take into
account the expansion of an unconfined wave function. As $Q$ is related to the typical gradient
of the wave function amplitude it follows that as the function becomes smeared over time and the gradient becomes small the quantum correction becomes negligible. To put in quantitative terms:
\beq
\vec F_Q = - \vec \nabla Q \simeq \frac{\hbar^2}{2 m_e L_R^3}, \qquad
  L_R \simeq \frac{R}{|\vec \nabla R|}
\label{quantcorrect}
\enq
in which $L_R$ is the typical length of the amplitudes gradient. Thus:
 \beq
 |F_Q|  << |F_L| \Rightarrow L_R >> L_{Rc} = \left(\frac{\hbar^2}{2 m_e F_L}\right)^\frac{1}{3}.
\label{quantcorrect2}
\enq
in which $\vec F_L = e  (\vec E + \vec v \times \vec B)$ is the classical Lorentz force \cite{BEP}. For the current application in which a free electron transverses a macroscopic length this term will certainly be negligible. However, for a confined electron this term cannot be neglected as we show in a following section describing the hydrogen atom.

\section {Pauli's electron}

\Sc's quantum mechanics is limited to the description of spin less particles. Indeed the need for
spin became necessary as \SE could not account for the result of the Stern Gerlach experiments, predicting a single spot instead of the two spots obtained for hydrogen atoms. Thus Pauli introduced his equation for a non-relativistic particle with spin is given by:
\beq
i \hbar \dot{\psi} = \hat{H}_P \psi, \qquad \hat{H}_P = -\frac{\hbar^2}{2 m_e}[\vec \nabla-\frac{ie}{\hbar }\vec A]^2 + \mu \vec B \cdot \vec \sigma + e \Phi
= \hat{H}_S ~ I + \mu \vec B \cdot \vec \sigma
 \label{Pauli}
 \enq
$\psi$ here is a two dimensional complex column vector (also denoted as spinor), $\hat{H}_P$ is a two dimensional hermitian operator matrix, $\mu$ is the magnetic moment of the particle, and $I$ is
a two dimensional unit matrix. $\vec \sigma$ is a vector of two dimensional Pauli matrices which can be represented as follows:
\beq
\sigma_{1} = \left( \begin{array}{cc} 0 & 1 \\ 1 & 0 \end{array} \right), \qquad
\sigma_{2} = \left( \begin{array}{cc} 0 & -i \\ i & 0 \end{array} \right), \qquad
\sigma_{3} = \left( \begin{array}{cc} 1 & 0 \\ 0 & -1 \end{array} \right).
\label{sigma}
\enq
The ad-hoc nature of this equation was later amended as it became clear that this is the non relativistic limit of the relativistic Dirac equation.
A spinor $\psi$ satisfying \ern{Pauli} must also satisfy a continuity equation of the form:
\beq
\frac{\partial{\rho_p}}{\partial t} + \vec \nabla \cdot  \vec j  = 0.
\label{massconp}
\enq
In the above:
\beq
\rho_p =  \psi^\dagger \psi, \qquad \vec j =  \frac{\hbar}{2 m_e i} [\psi^\dagger \vec \nabla \psi -(\vec \nabla \psi^\dagger) \psi]
- \frac{e}{m_e} \vec{A} \rho_p.
\label{pcd}
\enq
The symbol $\psi^\dagger$ represents a row spinor (the transpose) whose components are equal to the complex conjugate of the column spinor $\psi$. Comparing the standard continuity equation to \ern{massconp} suggests the definition of a velocity field as follows \cite{Holland}:
\beq
\vec v = \frac{\vec j}{\rho_p}= \frac{\hbar}{2 m_e i\rho_p} [\psi^\dagger \vec \nabla \psi -(\vec \nabla \psi^\dagger) \psi] - \frac{e}{m_e}  \vec{A}.
\label{pv}
\enq
 Holland \cite{Holland} has suggested the following representation of the spinor:
 \beq
\psi = R e^{i\frac{\chi}{2}} \left( \begin{array}{c} \cos \left(\frac{\theta}{2}\right) e^{i\frac{\phi}{2}} \\
i \sin \left(\frac{\theta}{2}\right) e^{-i\frac{\phi}{2}}  \end{array} \right) \equiv
\left( \begin{array}{c} \psi_{\uparrow} \\ \psi_{\downarrow} \end{array} \right).
\label{psiH}
\enq
In terms of this representation the density is given as:
\beq
R^2 = \psi^\dagger \psi = \rho_p  \Rightarrow R= \sqrt{\rho_p}.
\label{rhopsi}
\enq
The mass density is given as:
\beq
\hat \rho = m_e \psi^\dagger \psi =m_e R^2 = m_e \rho_p .
\label{rhom}
\enq
The probability amplitudes for spin up and spin down electrons are given by:
\beq
a_{\uparrow} = \left | \psi_{\uparrow} \right| = R \left | \cos \frac{\theta}{2} \right|, \qquad
a_{\downarrow} = \left | \psi_{\downarrow} \right| = R \left | \sin \frac{\theta}{2} \right|
\label{probamp}
\enq
Let us now look at the expectation value of the spin:
\beq
<\frac{\hbar}{2} \vec \sigma> = \frac{\hbar}{2}\int \psi^\dagger \vec \sigma \psi d^3 x =
 \frac{\hbar}{2}\int \left(\frac{\psi^\dagger \vec \sigma \psi}{\rho_p}\right) \rho_p d^3 x
\label{spinex}
\enq
The spin density can be calculated using the representation given in \ern{psiH} as:
\beq
\hat s \equiv \frac{\psi^\dagger \vec \sigma \psi}{\rho_p} = (\sin \theta \sin \phi, \sin \theta \cos \phi, \cos \theta), \qquad |\hat s| = \sqrt{\hat s \cdot \hat s} = 1.
\label{spinden}
\enq
This gives an easy physical interpretation to the variables $\theta,\phi$ as angles which describe the projection
of the spin density on the axes. $\theta$ is the elevation angle of the spin density vector and $\phi$ is
the azimuthal angle of the same. The velocity field can now be calculated by inserting $\psi$ given in \ern{psiH} into \ern{pv}:
\beq
\vec v = \frac{\hbar}{2 m_e} (\vec \nabla \chi + \cos \theta \vec \nabla \phi) - \frac{e}{m_e} \vec{A}.
\label{pv2}
\enq
We are now in a position to calculate the material derivative of the velocity and obtain the equation of motion for a particle with  (\cite{Holland} p. 393 equation (9.3.19)):
\beq
\frac{d \vec v}{d t}  = - \vec \nabla ( \frac{Q}{m_e})-\left(\frac{\hbar}{2 m_e}\right)^2
\frac{1}{\rho_p} \partial_k(\rho_p \vec \nabla \hat s_j \partial_k \hat s_j)
+\frac{e}{m_e}  (\vec E + \vec v \times \vec B) - \frac{\mu}{m_e} (\vec \nabla B_j) \hat s_j.
\label{EulerP}
\enq
The Pauli equation of motion differs from the classical equation motion and the Schr\"{o}dinger's equation of motion. In addition to the \Sc~ quantum force correction we have an additional spin quantum force correction:
\beq
\vec F_{QS} \equiv  -\frac{\hbar^2}{4 m_e}
\frac{1}{\rho_p} \partial_k(\rho_p \vec \nabla \hat s_j \partial_k \hat s_j)
= -\frac{\hbar^2}{4 m_e} \left[
\partial_k( \vec \nabla \hat s_j \partial_k \hat s_j)
+ \frac{\partial_k \rho_p }{\rho_p} \vec \nabla \hat s_j \partial_k \hat s_j
\right]
\label{FQS}
\enq
as well as a term characterizing the interaction of the spin with a gradient of the magnetic field,
which is the Stern-Gerlach term.
\beq
\vec F_{grad B S} \equiv  - \mu (\vec \nabla B_j) s_j
\label{FgrBS}
\enq
As both the upper and lower spin components of the wave function are expanding in free space the gradients which appear in $\vec F_{QS}$ will tend to diminish for any macroscopic scale making this
force negligible. To estimate the condition quantitatively we introduce the typical spin length:
\beq
L_s = {\rm min}_{~i\in\{1,2,3\}} ~|\vec \nabla \hat s_i|^{-1}
\label{Ls}
\enq
Using the above definition we may estimate the spin quantum force:
\beq
 F_{QS} \approx \frac{\hbar^2}{4 m} [\frac{1}{L_s^3} + \frac{1}{L_s^2 L_R}]
=  \frac{\hbar^2}{4 m L_s^2} [\frac{1}{L_s} + \frac{1}{L_R}]
\label{FQSest}
\enq
this suggested the definition of the hybrid typical length:
\beq
 L_{sR} = [\frac{1}{L_s} + \frac{1}{L_R}]^{-1} = \left\{
                                                          \begin{array}{cc}
                                                            L_s & L_s \ll L_R \\
                                                            L_R & L_R \ll L_s \\
                                                          \end{array}
                                                        \right. .
\label{LSR}
\enq
In terms of this typical length we may write:
\beq
 F_{QS} \approx \frac{\hbar^2}{4 m_e L_s^2  L_{sR}}
\label{FQSest2}
\enq
Thus the conditions for a classical trajectory become:
 \beq
 F_{QS} \ll F_L  \Rightarrow L_s^2  L_{sR} \gg \frac{\hbar^2}{4 m_e F_L}  \Rightarrow
 L_s \gg \left\{
     \begin{array}{cc}
       \left(\frac{\hbar^2}{4 m_e F_L}\right)^{\frac{1}{3}}  & L_s \ll L_R \\
        \left(\frac{\hbar^2}{4 m_e F_L L_{R}}\right)^{\frac{1}{2}} & L_R \ll L_s \\
        \end{array}
    \right. .
\label{FQSest3}
\enq
Another important equation derived from \ern{Pauli} is the equation of motion for the spin orientation vector (\cite{Holland} p. 392 equation (9.3.16)):
\beq
\frac{d \hat s}{d t}  = \frac{2 \mu}{\hbar} \vec B_{eff} \times \hat s,
\qquad
\vec B_{eff} = \vec B  - \frac{\hbar^2}{4 \mu m_e R^2} \partial_i (\rho \partial_i  \hat s)
\label{Spinequ}
\enq
The quantum correction to the magnetic field explains \cite{Holland} why a spin picks up the orientation of the field in a Stern-Gerlach experiment instead of precessing around it
as a classical magnetic dipole would.

In the free electron scenario the only quantum term that might have a significance is the Stern-Gerlach term given in \ern{FgrBS}, however, it is well known that this term is negligible
with respect to the Lorentz classical term, which is why Stern-Gerlach experiments are performed using natural particles. Thus as far as free electrons based relativistic engines are concerned,
a classical analysis will suffice, this is not the case for a confined electron as we discuss in the next section.

\section{The Hydrogen atom}

The Hydrogen atom is one of the simplest quantum mechanical systems and was discussed in \cite{nano}, we use the same notation as in \cite{nano} and will not redefine the notation here.

The associated velocity field of an eigen function is determined from its phase (equation (44) in \cite{nano}) is:
\beq
\vec v_S = \frac{\hbar}{m_e} \vec \nabla \phi - \frac{e}{m_e} \vec A
= \frac{m \hbar}{m_e r \sin \theta'} \hat \varphi,  \qquad \vec A = 0,
\label{vSH}
\enq
which is azimuthal, thus for every eigenstate the electron is circulating some axis (the $z$ axis which is arbitrarily defined). The speed of the electron in the hydrogen atom is thus:
\beq
v_S = \frac{m \hbar}{m_e r |\sin \theta'|}.
\label{sSH}
\enq
The speed will vanish for every eigenstate with a magnetic quantum number $m=0$ including for the ground state. However, for every other magnetic quantum number the velocity field is singular both in the proton at $r=0$ and on the north and south poles $\theta' =0,\pi$. Regrading the singularity at $r=0$ this is not problem from a physical point of view as one can expect a different potential from the Coulomb potential inside the proton which is not a point particle. However, with regard to the south and north poles infinite velocities, this indicates a difficulty in the Hydrogen atom classical description in which relativistic considerations which enforce speeds smaller than the velocity of light $c$ will be part of the solution. The static electron implies according to
\ern{EulerS} that the force is zero. This is indeed the case, one can calculate the quantum potential for every state by using \ern{qupo}, however, for Hydrogen eigenstates it will be easier to use \ern{nrhje} and substitute the phase from equation (44) of \cite{nano}, this gives the expression:
\beq
Q_{nm} = E_n - \frac{m^2 \hbar^2}{2 m_e r^2 \sin^2 \theta'}+ e \Phi
= E_n - \frac{m^2 \hbar^2}{2 m_e r^2 \sin^2 \theta'}+ \frac{k e^2}{r}
\label{Qnm}
\enq
this can be verified by direct substitution of eigenstates in \ern{qupo}.
It is easy to see that for $m=0$ the total force vanishes:
\beq
 \vec \nabla (Q_{nm}- e \Phi) = \vec \nabla E_n = 0
 \label{Qnm=0}
\enq
We are now in a position to calculate the current density given in \ern{scd}
\beq
\vec J   = \frac{\hbar}{m_e}  \rho \vec{\nabla} \phi =- m \frac{e \hbar}{m_e} |\psi_{nlm}|^2
\frac{\hat \varphi}{r \sin \theta'}
 \label{currdenhyd}
 \enq
We notice that the current density is linear in the magnetic number $m$, in particular if
$m=0$ there is no current density and thus no relativistic motor effect. We conclude that for an isolated hydrogen in the ground state $n=1,l=0,m=0$ there is no relativistic motor effect. But also in excited states in which the current density does not necessarily vanish there will be no relativistic motor effect if the potential acting on the electron is spherically or cylindrically symmetric as is evident from equation (3) of \cite{nano}.

 How can we use an hydrogen atom as a component in a relativistic motor despite
the fact that it is useless  either in the ground state or in an excited state? In the following
section we will suggest an approach in which the electron is not in an energy eigen state but in a superposition of states.

To understand the order of magnitude of the relativistic motor effect using a hydrogen atom
one is referred to \cite{nano}.

\section{A simple wave packet}

Let us assume an idealized wave packet of the form:
\beq
\psi  = A e^{i k' x}, \qquad A = \left\{ \begin{array}{cc}
                                           \sqrt{\tilde{\rho}_c} & r<R_{max} \\
                                           0 & r \ge R_{max} \\
                                         \end{array} \right.
 \label{psiwp}
 \enq
 in the above $k'$ and $\tilde{\rho}_c$ are constants. As the wave function must be normalized
 it follows  that $\tilde{\rho}_c$ must take the following value:
\beq
\tilde{\rho}_c = \frac{3}{4 \pi} R_{max}^{-3}
 \label{tilrhoc}
 \enq
 hence this wave function has a linear phase and a uniform amplitude which is confined inside a sphere of radius $R_{max}$. It is certainly not an eigen state of the Hydrogen atom Hamiltonian,
 the preparation of such a state will require a suitable electromagnetic field
  which will be discussed below. We have analyzed the properties of this wave function in \cite{nano} and will not repeat the analysis here.

The purpose of the wave function engineering is to achieve a wave function that will produce a stable linear momentum over macroscopic durations. This implies according to equation (3) of \cite{nano} and \ern{scd} that we need to achieve a constant wave packet amplitude and constant phase gradient affected by
a constant vector potential. A constant phase gradient does not imply a constant phase, in fact
we may write the phase in the form:
\beq
\phi = \phi_s (\vec x) -  \phi_t (t)
\label{phipatr}
\enq
for a time independent amplitude it follows from \ern{psi1}:
\beq
\frac{\partial \psi}{\partial t} = - i \psi  \frac{\partial \phi_t}{\partial t}
\label{tindpsi}
\enq
Defining:
\beq
E(t) = \hbar \frac{\partial \phi_t}{\partial t}
\label{Et}
\enq
which is a time dependent function with units of energy, \Sc~ \ern{Seq} implies that:
\beq
\hat{H}_S \psi = E(t) \psi
\label{SEq}
\enq
Thus to achieve such a condition $\psi$ must be an eigen function of some Hamiltonian $\hat{H}_S$ with a possibly time dependent eigenvalue $E(t)$. A Hamiltonian can be constructed by introducing
suitable electromagnetic fields into the physical system. For example let us consider the somewhat artificial wave packet described in \ern{psiwp} which we now augment with a time dependent phase:
\beq
\psi  = Am e^{i (k' x- \phi_t (t)}, \qquad Am = \left\{ \begin{array}{cc}
                                           \sqrt{\tilde{\rho}_c} & r<R_{max} \\
                                           0 & r \ge R_{max} \\
                                         \end{array} \right.
\label{psiwp22}
\enq
We shall now plug the above expression into \ern{Seq} and ignore the nonphysical derivatives connected to the fact that the above oversimplified wave packet is not smooth at $r=R_{max}$, it follows that:
\beq
E(t) = \frac{\hbar^2 k'^2}{2 m_e} - e \Phi - \frac{e \hbar k'}{m_e} A_x + \frac{e^2 A^2}{2 m_e},
\label{Eequ}
\enq
in the above we took advantage of the gauge freedom and assumed a Coulomb gauge $\vec \nabla \cdot \vec A = 0 $ which is of course not physically restrictive. This allows two types of solutions. In one case we assume $A_x = 0$, that is we assume that there is no vector potential component in the direction of motion of the wave packet. Denoting the perpendicular vector potential as $\vec A_\bot = A_y \hat y +  A_z \hat z$ it follows that:
\beq
A_\bot = \pm \frac{\sqrt{2 m_e}}{e} \sqrt{E(t) + e \Phi - \frac{\hbar^2 k'^2}{2 m_e}}.
\label{Aperp}
\enq
If, however, $A_x \neq 0$ it follows that:
\beq
A_x = \frac{1}{e} \left(\hbar k \pm \sqrt{2 m_e E(t) - e^2 A_\bot^2 + 2 m_e e \Phi}\right).
\label{Ax}
\enq

\section{Discussion}

The main results of this paper are the possibility of implementing a relativistic motor in the atomic and nano scales. It is shown that two approaches are possible. In one case we consider
free propagating electrons which moves nevertheless in proximity to the nucleus but have enough energy not to be captured by the nucleus, we also consider the case of confined electrons.

Free electrons are classical and quantum forces are shown to be negligible due to the phenomena of wave packet spreading, thus a relativistic engine based on free electrons is analyzed classically.

For confined electrons quantum effects are important. Unfortunately it is shown that a Hydrogen atom whether in a ground or excited state does not produce any momentum according to the relativistic motor equation. We study the case in which an electron is put in a wave packet state which is an eigen state of an unspecified Hamiltonian. The Electromagnetic field to generate such a Hamiltonian are calculated.

\section{Conclusion}

The requirement to construct an engine suitable for interplanetary travel which is based on the rocket effect, entails an enormous supply of fuel to be carried with the vehicle. Basically most of the spacecraft should be fuel. An alternative is thus suggested based on the relativistic motor effect, in which no fuel is needed.

Despite the theoretical possibility to construct a working relativistic motor suitable for space craft which are intended for interplanetary travel, in practice this will not be a trivial task and will involve the generation of a highly localized wave packet or alternatively a very narrow electron beam. Thus in a study which is not a merely preliminary as this one, a more realistic wave packet should be considered and the sources of the electromagnetic field needed to achieve this goal need to be specified.

Additional directions for future studies which are arise from this paper include:
\begin{enumerate}
  \item The analysis of a relativistic motor of which its components move also at relativistic speeds and not just the electromagnetic signals transmitted between the components. The need for this arises as the electron studied in the current paper may move at relativistic speeds.
  \item For the same reason an analysis of the relativistic motor in the frame work of a Dirac theory is required. The Schr\"{o}dinger equation and even the Pauli equation are not appropriate for the study of an electron at relativistic speeds.
  \end{enumerate}

%%%%%%%%%%%%%%%%%%%%%%%%%%%%%%%%%%%%%%%%%%
\end{paracol}
%%%%%%%%%%%%%%%%%%%%%%%%%%%%%%%%%%%%%%%%%%
% To add notes in main text, please use \endnote{} and un-comment the codes below.
%\begin{adjustwidth}{-5.0cm}{0cm}
%\printendnotes[custom]
%\end{adjustwidth}
%%%%%%%%%%%%%%%%%%%%%%%%%%%%%%%%%%%%%%%%%%
\reftitle{References}


%\externalbibliography{yes}
%\bibliography{your_external_BibTeX_file}

%=====================================
% References, variant B: internal bibliography
%=====================================

\end{document}